\documentclass[usenatbib]{mn2e}
\usepackage{color} 

\usepackage{graphics}
\usepackage{epsfig}
\usepackage{natbib}

\begin{document}

\title[MgII absorbers around low mass galaxies]
{Large scale correlations in gas traced by MgII absorbers around low mass galaxies}

\author [G.Kauffmann] {Guinevere Kauffmann\thanks{E-mail: gamk@mpa-garching.mpg.de}\\
Max-Planck Institut f\"{u}r Astrophysik, 85741 Garching, Germany}

\maketitle

\begin{abstract} 
The physical origin of the large-scale conformity in the colours and
specific star formation rates of isolated low mass central galaxies and their
neighbours on scales in excess of 1 Mpc is still under debate.  One
possible scenario is that gas is heated over large scales by feedback from
active galactic nuclei (AGN), leading to coherent modulation of cooling and
star formation between well-separated galaxies. In this Letter, the metal
line absorption catalogue of Zhu \& M\'enard (2013) is used to  probe gas
out to large projected radii around a sample of a million  galaxies with
stellar masses $\sim 10^{10} M_{\odot}$ and  photometric redshifts in the
range $0.4<z<0.8$ selected from Sloan Digital Sky Survey imaging data.
This galaxy sample covers an effective volume of 2.2 Gpc$^3$.  A
statistically significant excess of MgII absorbers is present around the
red low mass galaxies compared to their blue counterparts out to projected
radii of 10 Mpc. In addition, the equivalent width  distribution function
of  MgII absorbers around low mass galaxies is shown to be strongly
affected by the presence of a nearby ($R_p<2$ Mpc) radio-loud AGN  out to
projected radii of 5 Mpc.
\end{abstract}


\section {Introduction}

The physical origin of the large-scale conformity in the colours and
specific star formation rates of isolated central galaxies and their
neighbours on scales in excess of 1 Mpc has been the subject of
considerable recent controversy in the literature. In a series of papers,
Kauffmann et al (2013) and Kauffmann (2015) argue in favour of a
``pre-heating''  scenario where gas is heated over large scales, leading to
coherent modulation of cooling and star formation between well-separated
galaxies.

The alternative explanation is that these large-scale correlations arise as
a consequence of the gravitational physics of non-linear clustering, for
example so-called ``assembly bias'' effects that cause the formation times
of neighbouring dark matter halos to be correlated over large scales (Gao,
Springel \& White 2005).  Hearin et al (2015) construct a simple scheme for
populating dark matter halos with galaxies in which the  age of the central
galaxy is matched to the age of its surrounding dark matter halo, and show
that strong conformity effects then arise naturally. Pahwa \& Parajape (2017)
present a model where conformity effects arise as a result of the correlation
between galaxy colour and dark matter halo concentration index.
Semi-analytic
models of galaxy formation, which aim to reproduce a wider variety of
observational data than simplistic models linking galaxy colour to
halo properties, 
(Kauffmann et al 1999; Croton et
al 2006; Guo et al 2011; Henriques et al 2015) do not, however, yield conformity
effects as strong as those measured from Sloan Digital Sky Survey data
(Kauffmann et al 2013; Lacerna et al 2017). Given the
disagreement between different theoretical analyses,  it is
 important to try and pin down the cause
of the effect in a model-independent way.

To help distinguish between the pre-heating and assembly bias scenarios,
Kauffmann (2015) searched for {\em relics} of energetic feedback events in
the neighbourhood of central galaxies with different specific SFRs.  A
significant excess of very high mass ($\log M_* > 11.3$) galaxies out to a
distance of 2.5 Mpc around low SFR/$M_*$ central galaxies was found when
compared to control samples of higher SFR/$M_*$ central galaxies with the
same stellar mass and redshift. Very massive galaxies in the neighbourhood
of low-SFR/$M_*$ galaxies were also found to have much higher probability
of hosting radio-loud active galactic nuclei (AGN). Kauffmann (2015)
claimed that these results lend credence to the hypothesis that AGN might
be responsible for heating gas over large spatial scales.  This conclusion
has recently been challenged  by Tinker et al (2017) and  Sin, Lilly \& Henriques (2017), who carry
out a re-analysis of the data set analyzed by Kauffmann (2015) and point
out that a large part of  the large-scale conformity signal originates from
a small number of central galaxies in the vicinity of a small number of groups and  
massive clusters. In addition, Berti et al (2017) have studied conformity effects 
at redshifts $0.2<z<1$ using galaxies selected from the 
PRism MUlti-object Survey (PRIMUS). Although conformity effects are detected,
they are weaker than those found in the SDSS analysis of Kauffmann et al (2013) and
are argued to be the result of large-scale tidal fields and 
assembly bias.  

In this Letter, I carry out a first set of direct tests of the gas pre-heating hypothesis.
Sample statistics are improved by expanding the
number of low mass central galaxies  by a factor of more than a hundred
over that used in Kauffmann et al (2013) using   
a sample of a million  galaxies with photometric redshifts in the
range $0.4<z<0.8$ and stellar masses $10< \log M_* < 10.5$ selected from 
Sloan Digital Sky Survey imaging data.  This sample covers an effective
volume of 2.2 Gpc$^3$. I  use a sample of MgII quasar absorption line
systems from the metal line absorption catalogue of Zhu \& M\'enard (2013)
to probe the gas out to large projected radii around these galaxies as a
function of their colour. I show that a statistically significant excess of
MgII absorbers is present around the red low mass galaxy population out to
projected radii of 10 Mpc. In addition, I show that the presence of a
nearby massive galaxy with a radio loud AGN strongly affects the equivalent
width distribution function of  MgII absorbers around low mass galaxies out
to projected radii of 5 Mpc.

\section {The Samples}

\subsection {Low mass photometrically-selected galaxy sample} I make use of
the photometric redshift probability distributions for SDSS galaxies using
the methodology described in Sheldon et al (2012). These authors used a
nearest-neighbor weighting algorithm developed  by Lima et al (2008) 
to derive the ensemble redshift distribution N(z), and
individual redshift probability distributions P(z) for galaxies with $r <
21.8$.  As part of this technique, weights were calculated for a set of
training galaxies with known redshifts such that their density distribution
in five-dimensional colour/magnitude space was proportional to that of the
photometry-only sample, producing a nearly fair sample in that space. P(z)
for individual objects were derived by using training-set objects from the
local colour/magnitude space around each photometric object.  The largest
spectroscopic training set galaxies at fainter magnitudes consists of
16,874 from four fields of the PRIMUS survey (PRIMUS; Coil et al. 2011).
PRIMUS obtains redshifts to a precision of less than 0.4 \% for galaxies
with $i< 22.5$. The catalogue is publicly available at
https://data.sdss.org/sas/dr13/eboss/photoObj/photoz-weight/.

I select all galaxies from this catalogue where the integrated probability
for the redshift of the galaxy to lie in the range $0.4 < z < 0.8$ is
greater than 0.85. This redshift range is chosen because it overlaps with
the CMASS sample of massive galaxies with spectroscopic redshifts (see next
subsection), as well as the MgII catalogue of Zhu \& M\'enard (2013).  This
results in a sample of 5,474,672 photometrically selected galaxies. I
derive an estimate of the stellar mass of each galaxy by comparing their
$g-r$, $r-i$ and $i-z$ colours to a set of templates generated using the
stellar population synthesis models of Bruzual \& Charlot (2003) (I assume
that the galaxy is located at redshift $z_{med}$, where $z_{med}$ is the
median of the redshift probability distribution $P(z)$).  975,540
photometrically selected galaxies with derived stellar masses in the range
$10<\log M_*<10.5$ are included in this low mass sub-sample of galaxies.

\subsection {Sample of massive galaxies and radio-loud AGN} The sample of
massive galaxies used in this work originates from the twelfth data release
(DR12) of the SDSS.  Spectra of about 1.5 million galaxies are available
from the Baryonic Oscillation Spectroscopic Survey (BOSS; Dawson et al
2013). I select a sub-sample of 937,079 galaxies with 
redshifts $0.4 < z < 0.8$  that meet the CMASS target
selection criterion, which was designed to select an approximately stellar
mass limited sample at $z > 0.4$. To find radio-loud galaxies, I  cross-match
the CMASS sample with the source catalog from the Faint Images of the
Radio Sky at Twenty-Centimeters (FIRST) survey carried out at the VLA
(Condon et al.  1998). The SDSS and FIRST positions are required to be
within 3 arc seconds of each other. There are 43,812 galaxies in the radio
loud galaxy sample.

\subsection {The MgII absorber sample} I utilize the catalogue of 52,243
MgII absorbers extracted from 142,012 quasar spectra from the DR12 (Zhu \&
Menard 2013). Of these, 11,035 MgII absorbers have redshifts that overlap
the CMASS galaxy sample described above.

\section {Results} The left and central panels of Figure 1 show plots of
the cumulative number of MgII absorbers (top), number of MgII absorbers
weighted by the equivalent width (EQW) of each absorber (middle), and the
number of massive CMASS galaxies (bottom) around galaxies in the
photometrically selected low mass sample.  The analysis is restricted to
photometrically selected galaxies with derived stellar masses in the range
$10<\log M_*<10.25$. The cumulative profiles are plotted from a radius of
100 kpc to a radius of 1.5 Mpc in the left panels, and from 1.5 Mpc to 10
Mpc in the central panels.  Solid red curves show results for galaxies in
the reddest colour bin, with $r-i$ colours (k-corrected to z=0.5) greater
than 0.8.  Solid blue curves show results for galaxies in the bluest colour
bin with $r-i$ colours (k-corrected to z=0.5) less  than 0.5. Dashed red
and blue curves show counts of MgII absorbers and massive galaxies for {\em
random catalogues} which contain the same numbers of objects as our
red/blue galaxy sub-samples, but where the sky positions have been
randomized. The {\em difference} between the blue and the red curves is
plotted in the right panels:  the solid lines show the difference between
the counts around the blue and red low mass galaxies, while the dashed
lines show the difference between the counts around the blue/red random
catalogue objects.

\begin{figure}
\includegraphics[width=91mm]{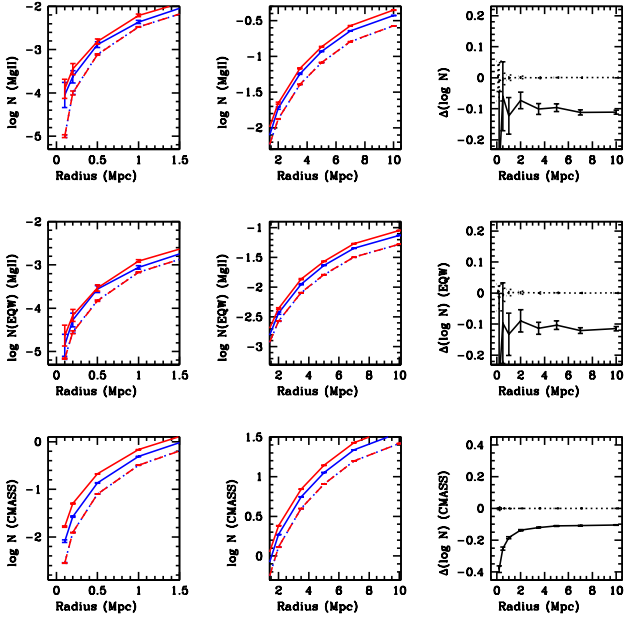}
\caption{ 
The cumulative number of MgII absorbers (top), number of MgII absorbers
weighted by the equivalent width (EQW) of each absorber (middle), and the
number of massive CMASS galaxies (bottom) around galaxies in the
photometrically selected sample of galaxies with
derived stellar masses in the range
$10<\log M_*<10.25$. The cumulative profiles are plotted from a radius of
100 kpc to a radius of 1.5 Mpc in the left panels, and from 1.5 Mpc to 10
Mpc in the central panels.  Solid red curves show results for galaxies in
the reddest colour bin, with $r-i$ colours (k-corrected to z=0.5) greater
than 0.8.  Solid blue curves show results for galaxies in the bluest colour
bin with $r-i$ colours (k-corrected to z=0.5) less  than 0.5. Dashed red
and blue curves show counts of MgII absorbers and massive galaxies for 
random catalogues. 
The {\em difference} between the blue and the red curves is
plotted in the right panels:  the solid lines show the difference between
the counts around the blue and red low mass galaxies, while the dashed
lines show the difference between the counts around the blue/red random
catalogue objects.
\label{models}}
\end{figure}

For both the red and the blue galaxy sub-samples, an  excess in MgII
absorbers and CMASS galaxies is detected over the random counts out to
projected radii of 10 Mpc.
The counts of CMASS galaxies are less noisy on
small scales and they rise steeply as a function of radius out to projected
radii of $\sim 2$ Mpc, reflecting the fact that at fixed stellar mass $M_*$,
red galaxies are likely to reside in more massive dark matter halos than
blue galaxies.
 More interesting is the fact that the counts of
both MgII absorbers and CMASS galaxies are different around red and blue
low mass galaxies out to the very largest scales. The difference in the
counts of CMASS galaxies remain roughly constant on scales from 2-10 Mpc
The same is seen for the MgII absorbers.
These results show that both the number of CMASS
galaxies and the gas traced by MgII is correlated with
the colour of low mass galaxies on  scales much larger than the
virial radii of their dark matter halos. 

One might ask whether the excess clustering on such large
scales is consistent with clustering statistics predicted in the standard
LCDM cosmology. Higher mass halos are more biased with respect to the underlying
dark matter density field than lower mass halos. In addition, there
are other effects that might come into play. Gao, White \& Jenkins (2005) studied the age dependence of
halo clustering on large scales (so-called halo assembly
bias)  using the  Millennium Simulation
carried out by the Virgo Consortium (Springel et al. 2005). Halos that form
at earlier epochs are more clustered than halos that form at later epochs, but
the effect decreases at high halo masses and is no longer present at
halo masses greater than $10^{13} h^{-1}$ Mpc. CMASS galaxies have been
shown to reside in halos with a mean mass of $5.2 \times 10^{13} M_{\odot}$
(Parejko et al 2013). Croton, Gao \& White (2007) examined halo assembly
bias effects on galaxy clustering as a function of colour using a ``shuffling''
technique and found that the effects disappeared for the galaxies with
luminosities comparable to those of the CMASS galaxies in our sample.

Further investigation of this issue is beyond the scope of this Letter.
Because the  low
mass galaxy sample is photometrically selected, it is not possible to split galaxies according to halo mass
reliably.
We note, however, that the  galaxy counts as a function of
colour are only an indirect probe of the physical effects of interest.
The next step is to look for evidence that the large-scale physical
properties of the gas around red/blue low mass galaxies are influenced by
the presence of a radio-loud AGN. Figure 2 shows the distribution of the
equivalent widths of the MgII absorbers around  the red and the blue
galaxies.  Results are plotted within apertures of projected radii 0.5,1, 2
and 5 Mpc.  As can be seen, the EQW distributions are identical for the two
sub-samples within all four apertures. Figure 3 shows the distribution of
the equivalent widths of the MgII absorbers around low mass galaxies with 0
(black lines), 1 (red lines) and 2-4 (magenta lines) CMASS galaxies located
within a projected radius of 2 Mpc. Results are shown for the same four
apertures as in Figure 1. Here, small but statistically significant
differences are apparent between the equivalent width distributions around
low mass galaxies with no massive neighbours and and around galaxies with 1
or more massive neighbours. In particular, there are fewer very low EQW
systems if a massive neighbour is present, but  more systems of
intermediate EQW. Finally, Figure 4 shows the distribution of the
equivalent widths of the MgII absorbers around low mass galaxies with 0
(black lines), 1 (red lines) and 2-4 (magenta lines) radio-loud CMASS
galaxies within a projected distance of 2 Mpc. In this plot,  very striking
differences are seen between the EQW distributions around galaxies with no
radio-loud neighbour and those with 1 or more such AGN in their vicinity,
which persists out to projected radii of 5 Mpc. Interestingly, the EQW
distribution around galaxies with a radio-loud neighbour appears to be
bimodal and more narrowly peaked, i.e. both low and high EQW absorbers
appear to be missing from the distribution.

We note that Best et al (2005) showed that  radio-loud AGN fraction  
is around 10\% in galaxies with stellar masses comparable to those in the CMASS
sample. All massive galaxies are believed to go through {\em cycles} of
radio jet production. Inverse-compton  ``ghosts'' of past radio jet interaction
with the surrounding hot medium are frequently found in galaxy groups and clusters
(e.g. Mocz, Fabian \& Blundell 2010). If radio-loud AGN are
responsible for the large-scale effects presented in this Letter, it is thus not 
unexpected that they be seen in slightly weaker form by simply requiring the
presence of a massive companion galaxy. 

\begin{figure}
\includegraphics[width=88mm]{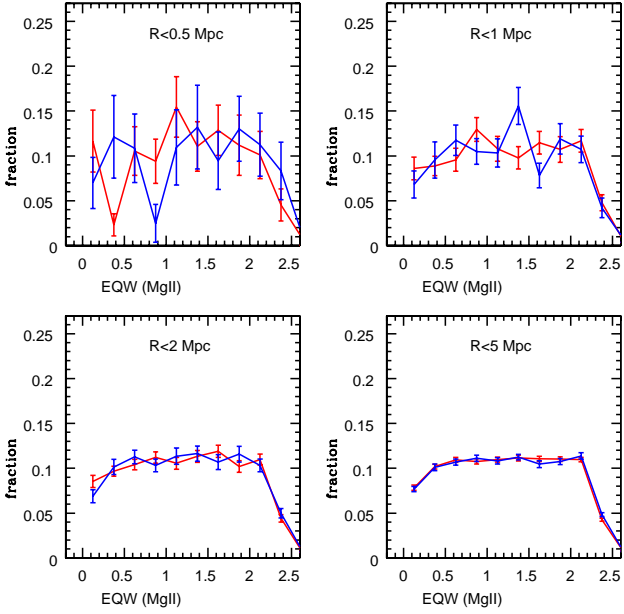}
\caption{ Thee distribution of the
equivalent widths of the MgII absorbers around  the red (red lines) and the blue (blue lines)
low mass galaxy sub-samples.  Results are plotted within apertures of projected radii 0.5,1, 2
and 5 Mpc. 
\label{models}}
\end{figure}

\begin{figure}
\includegraphics[width=88mm]{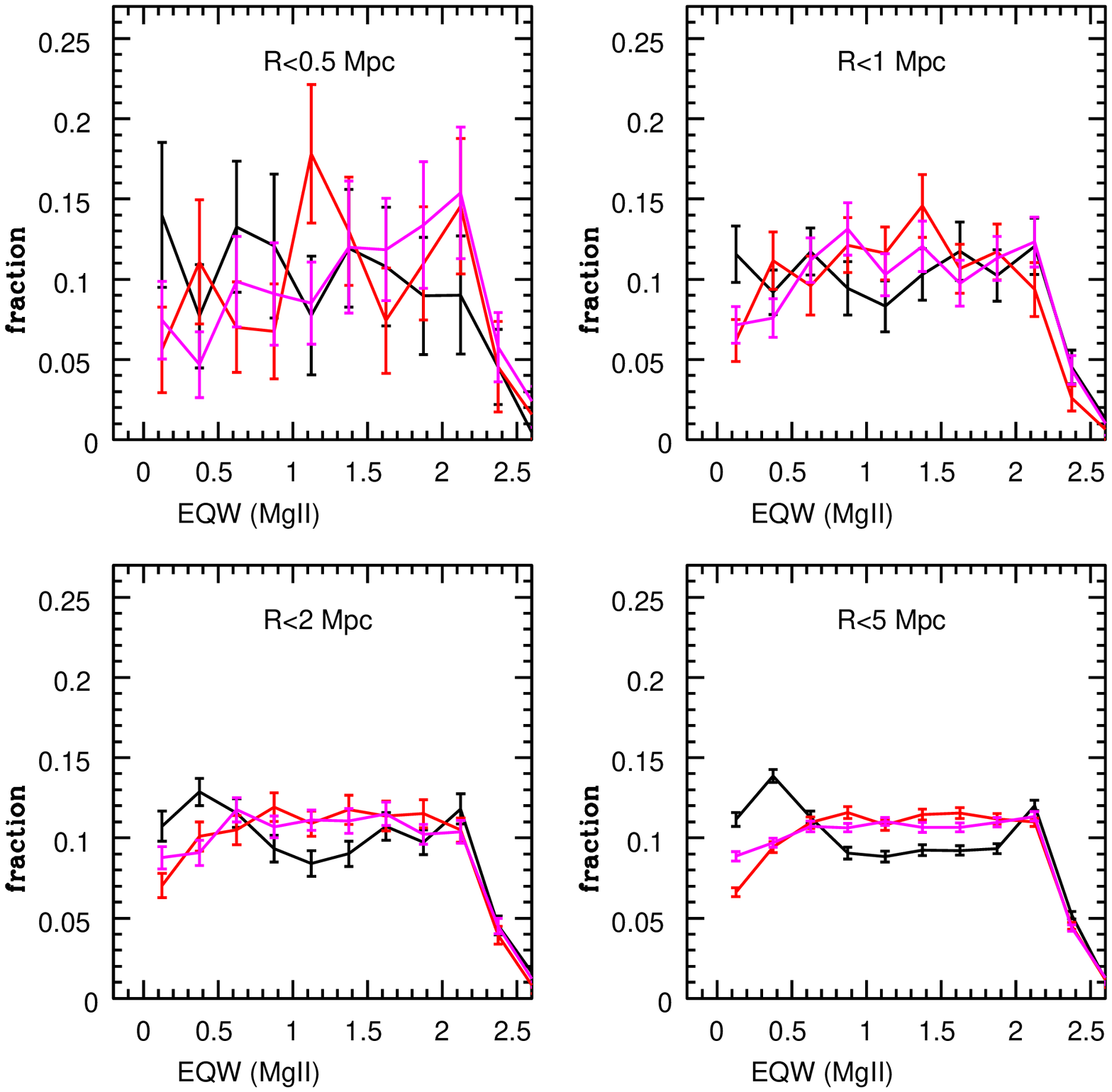}
\caption{ Thee distribution of the
equivalent widths of the MgII absorbers around  
low mass galaxies with 0
(black lines), 1 (red lines) and 2-4 (magenta lines) CMASS galaxies located
within a projected radius of 2 Mpc.
Results are plotted within apertures of projected radii 0.5,1, 2
and 5 Mpc. 
\label{models}}
\end{figure}

\begin{figure}
\includegraphics[width=88mm]{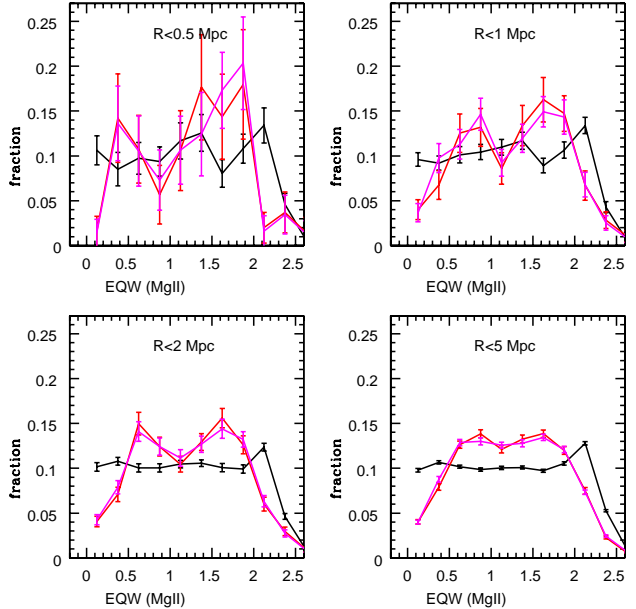}
\caption{ Thee distribution of the
equivalent widths of the MgII absorbers around  
low mass galaxies with 0
(black lines), 1 (red lines) and 2-4 (magenta lines) radio-loud CMASS galaxies located
within a projected radius of 2 Mpc.
Results are plotted within apertures of projected radii 0.5,1, 2
and 5 Mpc. 
\label{models}}
\end{figure}

\section {Summary and discussion} In this Letter, I present evidence that
some of the properties of the gas around low mass galaxies are coherent
over large spatial scales by showing that differences in the abundance of
MgII absorption lines systems around red and blue galaxies with stellar
masses $\sim 10^{10} M_{\odot}$ persist out to projected radii of 10 Mpc. I
present evidence that radio-loud AGN may be responsible for this
large-scale modulation by showing that the EQW distrbution of MgII systems
changes significantly if radio galaxy is found within 2 Mpc.  The change in
EQW distribution persists out to projected radii of 5 Mpc.

In order to make a direct link between these results and the question of
why the specific star formation rates of galaxies exhibit conformity over
large spatial scales, it would be helpful to be able to measure actual
physical properties of the gas, such as temperature and density. Gas in the
halos around galaxies is believed have complex structure and to exist in a
multi-phase configuration. This means that it is not simple or straightforward
to predict  MgII
equivalent width statistics for comparison with observations. It is still not
understood, for example, how numerical resolution influences predictions of
the detailed structure of the intergalactic gas in cosmological hydrodynamical 
simulations.  In
future, deep X-ray observations of high temperature gas, combined with
higher signal-to-noise quasar spectra that probe a variety of ionization
states will yield more insight into the scale dependence of the heating and
cooling processes occurring in the circumgalactic gas.


\end{document}